# Multi-user quantum private comparison with scattered preparation and one-way convergent transmission of quantum states

Tian-Yu Ye*, Zhao-Xu Ji

College of Information & Electronic Engineering, Zhejiang Gongshang University, Hangzhou 310018, P.R.China
*E-mail：happyyty@aliyun.com

**Abstract:** Quantum private comparison (QPC) aims to accomplish the equality comparison of the secrets from different users without disclosing their genuine contents by using the principles of quantum mechanics. In this paper, we summarize eight modes of quantum state preparation and transmission existing in current QPC protocols first. Then, by using the mode of scattered preparation and one-way convergent transmission, we construct a new multi-user quantum private comparison (MQPC) protocol with two-particle maximally entangled states, which can accomplish arbitrary pair's comparison of equality among $K$ users within one execution. Analysis turns out that its output correctness and its security against both the outside attack and the participant attack are guaranteed. The proposed MQPC protocol can be implemented with current technologies. It can be concluded that the mode of scattered preparation and one-way convergent transmission of quantum states is beneficial to designing the MQPC protocol which can accomplish arbitrary pair's comparison of equality among $K$ users within one execution.

## 1  Introduction

With the rapid development of quantum information processing in recent years, quantum cryptography, which was first invented by Bennett and Brassard [1] in 1984 with the concept of quantum key distribution (QKD), has attracted more and more attention. Up to date, quantum cryptography can be classified into several famous branches, such as QKD [1-8], quantum secret sharing (QSS) [9-13], quantum secure direct communication (QSDC) [14-28], quantum private query (QPQ) [29-33], *etc*.

Secure multi-user computation (SMC) is an interesting topic in classical cryptography. It is well known that the security of classical SMC relies on the computation complexity, which is susceptible to the powerful ability of quantum computation. In order to overcome its shortcoming on security, classical SMC has been generalized into its counterpart in the realm of quantum mechanics. Accordingly, quantum secure multi-user computation (QSMC) was derived.

As an important branch of QSMC, quantum private comparison (QPC), which was first proposed by Yang and Wen [34] in 2009 with Einstein-Podolsky-Rosen (EPR) pairs, aims to accomplish the equality comparison of the secrets from different users without disclosing their genuine contents based on the principles of quantum mechanics. However, since it is impossible to design a secure equality function in a two-party scenario [35], some additional assumptions, for example, a third party (TP), are always required in QPC. Up to now, numerous QPC protocols employing a TP [34,36-64] have been constructed with different quantum states and quantum technologies.

As far as the role of TP in QPC is concerned, there are three different models, which have been explained in detail by Zhang and Zhang [47]. (1) TP is honest. TP helps complete the protocol successfully and does not try to steal users' secrets. In this situation, the users only need to send their private secrets to TP through one-time-pad encryption, then TP compares the decrypted secrets and announces the comparison result. However, this situation is too perfect to be practical in reality. (2) TP is dishonest. He may refuse to participate, forge intermediate computations, stop the protocol before it is normally finished, or try to obtain the users' secrets. In this situation, the users cannot trust TP at all. This situation equals to the two-party scenario, whose insecurity has been validated by Lo in Ref.[35]. (3) TP is semi-honest, which implies two different definitions. The first one, introduced by Chen *et al.* [36] for the first time, is that TP executes the protocol loyally and keeps a record of all its intermediate computations and might try to steal the users' private secrets from the record, but he cannot be corrupted by the adversary including the dishonest user. However, Yang *et al.* [46] first pointed out definitely that this definition is unreasonable, and the reasonable one is that TP is allowed to try to steal the users' private secrets with any possible means but cannot be corrupted by the adversary including the dishonest user. Up to now, the second definition of a semi-honest TP has been widely accepted as the best assumption for TP in the realm of QPC.

Suppose that there are $K$ users, who want to know whether their individual secret is equal to either of the other $K-1$ secrets or not. If the two-user QPC protocol is used to solve this multi-user problem of equality comparison, it has to be executed with $(K-1) \sim K(K-1)/2$ times, which results in the reduction of efficiency. Under this circumstance, the first multi-user quantum private comparison (MQPC) protocol was suggested by Chang et al. [51] in 2013 with $n$-particle GHZ class states, which can accomplish arbitrary pair's comparison of equality among $K$ users within one execution. Subsequently, the MQPC protocols based on multi-level quantum system [62,63] and based on the entanglement swapping of two-level Bell entangled states [64] were designed, respectively. However, up to date, the number of MQPC protocols is still few. The research on MQPC should be accelerated.

In this paper, we first investigate the modes of quantum state preparation and transmission in current QPC protocols carefully, then utilize the mode of scattered preparation and one-way convergent transmission to design a special kind of MQPC protocol. It turns out that this mode is beneficial to designing the MQPC protocol which can accomplish arbitrary pair's comparison of equality among $K$ users within one execution.

The rest of this paper is structured as follows: the modes of quantum state preparation and transmission in current QPC protocols are summarized in Sec.2; a new MQPC protocol with two-particle maximally entangled states based on the mode of scattered preparation and one-way convergent transmission and its analysis are described in Sec.3; finally, discussion and conclusion are given in Sec.4.

## 2 Modes of quantum state preparation and transmission in current QPC protocols

After looking into the current QPC protocols [34,36-64] in depth, without considering the eavesdropping check processes, we summarize eight modes of quantum state preparation and transmission existing in these protocols as follows, which are also depicted in Fig.1.

**(1) Centralized preparation and circled transmission** (shown in Fig.1(a)). TP prepares quantum state $|\Phi\rangle$ as the quantum carrier first. Then, he transmits partial particles of his prepared quantum state, $|\Pi\rangle$, to the user $P_i$. After certain quantum operation, $P_i$ transmits $|\Pi\rangle$ to another user $P_j$. Finally, after certain quantum operation, $P_j$ transmits $|\Pi\rangle$ back to TP. In this mode, TP is the only person who prepares quantum state as the quantum carrier. Moreover, a part of particles from the prepared quantum state, $|\Pi\rangle$, is transmitted among TP, $P_i$ and $P_j$ in a circled way. The protocols in Refs.[38,42,49,63] adopt this kind of quantum state preparation and transmission mode.

**(2) Centralized preparation and one-way divergent transmission** (shown in Fig.1(b)). TP prepares quantum state $|\Phi\rangle$ as the quantum carrier first. Then, he transmits two parts of particles from his prepared quantum state, $|\Pi_1\rangle$ and $|\Pi_2\rangle$, to the users $P_i$ and $P_j$, respectively. In this mode, TP is the only person who prepares quantum state as the quantum carrier. Moreover, two parts of particles, $|\Pi_1\rangle$ and $|\Pi_2\rangle$, are transmitted from TP to $P_i$ and $P_j$ in a one-way divergent way. The protocols in Refs.[36,37,40,43-47,51,54-56,58-61,63] adopt this kind of quantum state preparation and transmission mode.

**(3) Centralized preparation and bidirectional divergent transmission** (shown in Fig.1(c)). TP prepares quantum state $|\Phi\rangle$ as the quantum carrier first. Then, he transmits two parts of particles from his prepared quantum state, $|\Pi_1\rangle$ and $|\Pi_2\rangle$, to the users $P_i$ and $P_j$, respectively. Finally, after certain quantum operation, $P_i$ and $P_j$ transmit $|\Pi_1\rangle$ and $|\Pi_2\rangle$ back to TP, respectively. In this mode, TP is the only person who prepares quantum state as the quantum carrier. Moreover, $|\Pi_1\rangle$ ($|\Pi_2\rangle$) is transmitted between TP and $P_i$ ($P_j$) forth and back. The protocols in Refs.[34,48,62] adopt this kind of quantum state preparation and transmission mode.

**(4) Scattered preparation and bidirectional transmission** (shown in Fig.1(d1&d2)). There are two cases for this mode. In the first case (shown in Fig.1(d1)), TP, $P_i$ and $P_j$ prepare quantum states $|\Phi\rangle$, $|\Phi_1\rangle$ and $|\Phi_2\rangle$ as the quantum carriers, respectively. Then, $P_i$ ($P_j$) transmits partial particles of her prepared quantum state, $|\Pi_1\rangle$ ($|\Pi_2\rangle$), to TP. Moreover, TP transmits two parts of particles from his prepared quantum state, $|\Pi_3\rangle$ and $|\Pi_4\rangle$, to $P_i$ and $P_j$, respectively. In this case, all of TP, $P_i$ and $P_j$ are the persons who prepare quantum state as the quantum carrier. Moreover, TP and $P_i$ ($P_j$) send particles to each other. The protocol in Ref.[39,64] adopts this kind of quantum state preparation and transmission mode. In the second case (shown in Fig.1(d2)), $P_i$ and $P_j$ prepare quantum states $|\Phi_1\rangle$ and $|\Phi_2\rangle$ as the quantum carriers, respectively. Then, $P_i$ ($P_j$) transmits partial particles of her prepared quantum state, $|\Pi_1\rangle$ ($|\Pi_2\rangle$), to $P_j$ ($P_i$). In this case, both $P_i$ and $P_j$ are the persons who prepare quantum state as the quantum carrier. Moreover, $P_i$ and $P_j$ send particles to each other. The protocol in Ref.[53] adopts this kind of quantum state preparation and transmission mode.

**(5) Scattered preparation and one-way convergent transmission** (shown in Fig.1(e)). $P_i$ and $P_j$ prepare quantum states $|\Phi_1\rangle$ and $|\Phi_2\rangle$ as the quantum carriers, respectively. Then, $P_i$ ($P_j$) transmits partial particles of her prepared quantum state, $|\Pi_1\rangle$ ($|\Pi_2\rangle$), to TP. In this mode, both $P_i$ and $P_j$ are the persons who prepare quantum state as the quantum carrier. Moreover, TP is the convergent destination of the particle transmissions from $P_i$ and $P_j$. The protocol in Ref.[41] adopts this kind of quantum state preparation and transmission mode.

**(6) Scattered preparation and hybrid transmission** (shown in Fig.1(f)). $P_i$ and $P_j$ prepare quantum states $|\Phi_1\rangle$ and $|\Phi_2\rangle$ as the quantum carriers, respectively. Then, $P_i$ ($P_j$) transmits partial particles of her prepared quantum state, $|\Pi_1\rangle$ ($|\Pi_2\rangle$), to TP. Moreover, $P_i$ ($P_j$) transmits partial particles of her prepared quantum state, $|\Pi_3\rangle$ ($|\Pi_4\rangle$), to $P_j$ ($P_i$). In this mode, both $P_i$ and $P_j$ are the persons who prepare quantum state as the quantum carrier. Moreover, there are both single-directional and bidirectional particle transmissions. The protocol in Ref.[50] adopts this kind of quantum state preparation and transmission mode.

**(7) Scattered preparation and one-way transmission** (shown in Fig.1(g)). $P_i$ and $P_j$ prepare quantum states $|\Phi_1\rangle$ and $|\Phi_2\rangle$ as the quantum carriers, respectively. Then, $P_i$ ($P_j$) transmits partial particles of her prepared quantum state, $|\Pi_1\rangle$ ($|\Pi_2\rangle$), to TP. Moreover, $P_i$ transmits partial particles of her prepared quantum state, $|\Pi_3\rangle$, to $P_j$. In this mode, both $P_i$ and $P_j$ are the persons who prepare quantum state as the quantum carrier. Moreover, all of the particle transmissions are single-directional. The protocol in Ref.[52] adopts this kind of quantum state preparation and transmission mode.

**(8) Centralized preparation and one-way transmission** (shown in Fig.1(h)). $P_i$ prepares quantum state $|\Phi\rangle$ as the quantum carrier first. Then, she transmits partial particles of her prepared quantum state, $|\Pi_1\rangle$, to $P_j$. In this mode, $P_i$ is the only person who prepares quantum state as the quantum carrier. Moreover, $|\Pi_1\rangle$ is transmitted from $P_i$ to $P_j$ in a single-directional way. The protocols in Ref.[57] adopt this kind of quantum state preparation and transmission mode.

In the following, we will use the mode of scattered preparation and one-way convergent transmission to design a special kind of MQPC protocol. It can be found out that this mode is beneficial to designing the MQPC protocol which can accomplish arbitrary pair's comparison of equality among $K$ users within one execution.

## 3 The proposed MQPC protocol
### 3.1 Protocol description

Assume that there are $K$ users, $P_1, P_2, \ldots, P_K$, where $P_i$ has a secret named $X^i$, $i = 1, 2, \ldots, K$. The binary representation of $X^i$ in $F_{2^N}$ is $\left(x_{L-1}^i, x_{L-2}^i, \ldots, x_0^i\right)$, where $x_l^i \in \{0,1\}$ and $l = 0, 1, \ldots, L-1$. They want to know whether every two different $X^i$'s are equal or not under the help of a semi-honest TP. Note that throughout this paper, we adopt Yang et al.[46]'s definition of the semi-honest TP, which is regarded to be the most reasonable. Here, in order to accomplish this goal, we construct a new MQPC protocol with two-particle maximally entangled states by using the mode of scattered preparation and one-way convergent transmission.

The Bell states defined in Eq.(1) are two-particle maximally entangled states, and form a complete orthogonal basis of four-dimensional Hilbert space. Without loss of generality, we use $|\phi^+\rangle$ as the quantum carrier here. Throughout this paper, the $\sigma_z$ basis means $\{|0\rangle, |1\rangle\}$.

$$|\phi^\pm\rangle = \frac{1}{\sqrt{2}}(|00\rangle \pm |11\rangle), |\psi^\pm\rangle = \frac{1}{\sqrt{2}}(|01\rangle \pm |10\rangle). \qquad (1)$$

The proposed MQPC protocol is described as follows.

***Preliminary:*** $P_i$ ($i = 1, 2, \ldots, K$) divides her binary representation of $X^i$ into $L$ groups, $G_{P_i}^1, G_{P_i}^2, \ldots, G_{P_i}^L$, where each group contains one binary bit of $X^i$. $P_i$ and $P_j$ ($i, j = 1, 2, \ldots, K$ and $i \neq j$) share one common key sequence, $K_{P_{ij}} = \left\{K_{P_{ij}}^1, K_{P_{ij}}^2, \ldots, K_{P_{ij}}^L\right\}$, with a QKD protocol [1-8] in advance, where $K_{P_{ij}}^k \in \{0,1\}$ and $k = 1, 2, \ldots, L$. Note that $K_{P_{ij}}$ is the same to $K_{P_{ji}}$.

***Step 1:*** $P_i$ ($i = 1, 2, \ldots, K$) prepares a quantum state sequence composed of $L$ $|\phi^+\rangle$, i.e., $S_{P_i} = \left[Q_{P_{i1}}^1 Q_{P_i T_1}^1, Q_{P_{i1}}^2 Q_{P_i T_1}^2, \ldots, Q_{P_{i1}}^L Q_{P_i T_1}^L\right]$. Here, the subscripts $P_{i1}, P_i T_1$ represent two particles of each $|\phi^+\rangle$, while the superscripts $1, 2, \ldots, L$ indicate the orders of $|\phi^+\rangle$ in $S_{P_i}$. Then, $P_i$ divides $S_{P_i}$ into two sequences. Concretely speaking, she picks particle $P_i T_1$ out from each $|\phi^+\rangle$ in $S_{P_i}$ to form an ordered sequence $S_{P_iT} = \left[ Q_{P_iT_1}^1, Q_{P_iT_1}^2, \ldots, Q_{P_iT_1}^L \right]$; the remaining particles in $S_{P_i}$ automatically compose another ordered sequence $S_{P_i}^{'} = \left[ Q_{P_{i1}}^1, Q_{P_{i1}}^2, \ldots, Q_{P_{i1}}^L \right]$.

**Step 2:** $P_i$ ($i=1,2,\ldots,K$) prepares one set of decoy photons, $D_i$, which is randomly chosen from the four states $\{|0\rangle, |1\rangle, |+\rangle, |-\rangle\}$, where $|\pm\rangle = \frac{1}{\sqrt{2}}(|0\rangle \pm |1\rangle)$. Then, $P_i$ randomly inserts $D_i$ into $S_{P_iT}$ to form a new sequence $S_{P_iT}^{'}$. Afterward, $P_i$ sends $S_{P_iT}^{'}$ to TP in the way of block transmission [14].

After confirming TP's receipt of $S_{P_iT}^{'}$, $P_i$ tells TP the positions and the preparation basis of decoy photons in $S_{P_iT}^{'}$. Then, TP measures the decoy photons in $S_{P_iT}^{'}$ with the basis $P_i$ told, and tells $P_i$ his measurement results. By comparing decoy photons' prepared initial states with TP's measurement results, $P_i$ can judge whether there is an eavesdropper or not during the transmission of $S_{P_iT}^{'}$. If $P_i$ discovers that there is an eavesdropper on line, the communication will be halted; otherwise, the communication will be continued.

**Step 3:** $P_i$ ($i=1,2,\ldots,K$) measures $Q_{P_{i1}}^k$ in $S_{P_i}^{'}$ with $\sigma_z$ basis. The measurement result of $Q_{P_{i1}}^k$ is denoted as $M_{P_i}^k$. If $M_{P_i}^k$ is $|0\rangle$, then $C_{P_i}^k$ is 0; and if $M_{P_i}^k$ is $|1\rangle$, then $C_{P_i}^k$ is 1. Here, $C_{P_i}^k$ is the classical one-bit code of $M_{P_i}^k$, and $k=1,2,\ldots,L$. Then, $P_i$ computes $R_{P_{ij}}^k = G_{P_i}^k \oplus C_{P_i}^k \oplus K_{P_{ij}}^k$, where $j=1,2,\ldots,K$ and $j \neq i$. Finally, $P_i$ publishes $R_{P_{ij}}$ to TP, where $R_{P_{ij}} = \left[ R_{P_{ij}}^1, R_{P_{ij}}^2, \cdots, R_{P_{ij}}^L \right]$. After receiving $R_{P_{ij}}$, TP keeps a record of $R_{P_{ij}}$ on his site.

**Step 4:** TP drops out the decoy photons in $S_{P_iT}^{'}$ ($i=1,2,\ldots,K$) to restore $S_{P_iT}$. Then, TP measures $Q_{P_iT_1}^k$ in $S_{P_iT}$ with $\sigma_z$ basis, where $k=1,2,\ldots,L$. The measurement result of $Q_{P_iT_1}^k$ is denoted as $M_{P_iT}^k$. TP keeps a record of $M_{P_iT}$ on his site, where $M_{P_iT} = \left[ M_{P_iT}^1, M_{P_iT}^2, \cdots, M_{P_iT}^L \right]$.

**Step 5:** For $k=1,2,\ldots,L$: According to the state of $M_{P_iT}^k M_{P_jT}^k$ ($i,j=1,2,\ldots,K$ and $i \neq j$), TP establishes a classical one-bit code $C_{T_{ij}}^k$. Concretely speaking, if $M_{P_iT}^k M_{P_jT}^k \in \{|0\rangle|0\rangle, |1\rangle|1\rangle\}$, TP sets $C_{T_{ij}}^k = 0$; and if $M_{P_iT}^k M_{P_jT}^k \in \{|0\rangle|1\rangle, |1\rangle|0\rangle\}$, TP sets $C_{T_{ij}}^k = 1$. Then, TP computes $R_{ij}^k = R_{P_{ij}}^k \oplus R_{P_{ji}}^k \oplus C_{T_{ij}}^k$. If $R_{ij}^k \neq 0$, TP concludes that $X^i \neq X^j$ and finishes the protocol immediately; otherwise, he sets $k = k+1$ and repeats from the beginning of this Step. If he finds out that $R_{ij}^k = 0$ for all $k$ in the end, he will conclude that $X^i = X^j$. Finally, TP tells $P_i$ and $P_j$ the comparison result of $X^i$ and $X^j$.

For clarity, the flow chart of this protocol is depicted in Fig.2, taking the comparison between $P_i$ and $P_j$ ($i,j=1,2,\ldots,K$ and $i \neq j$) for example. Apparently, each user individually prepares initial quantum states as quantum carriers and consistently transmits traveling particles to TP. It is equivalent to say that this protocol employs the mode of scattered preparation and one-way convergent transmission.

It is easy to know that within one execution, the proposed protocol can finish comparing the equality of $X^i$ and $X^j$, where $i,j=1,2,\ldots,K$ and $i \neq j$. In other words, the proposed protocol can accomplish arbitrary pair's comparison of equality among $K$ users within one execution. It is worthy emphasizing that in the proposed protocol, as for $P_i$, in order to judge whether her secret (i.e., $X^i$) is equal to any of the other $K-1$ secrets (i.e., $X^j$, $j=1,2,\ldots,K$ and $j \neq i$), she only needs to prepare $S_{P_i}$, send $S_{P_iT}^{'}$ to TP and measure $Q_{P_{i1}}^k$ in $S_{P_i}^{'}$ each for one time. In the meanwhile, as for TP, he only needs to measure $Q_{P_iT_1}^k$ in $S_{P_iT}$ for one time. It can be concluded that the mode of scattered preparation and one-way convergent transmission is beneficial to designing the MQPC protocol which can accomplish arbitrary pair's comparison of equality among $K$ users within one execution.

### 3.2 Analysis

We analyze the proposed protocol from the aspects concerning the output correctness and the security in this Section. The output correctness is analyzed in Subsec.3.2.1, while the security is analyzed in Subsec.3.2.2.

#### 3.2.1 Output correctness

Here, we take the comparison between $P_i$ and $P_j$ ($i,j=1,2,\ldots,K$ and $i \neq j$) for example to demonstrate the output correctness of the proposed protocol. In this situation, the relations among essential parameters $M_{P_i}^k$, $C_{P_i}^k$, $M_{P_iT}^k$, $M_{P_j}^k$, $C_{P_j}^k$, $M_{P_jT}^k$, $C_{T_{ij}}^k$ and $C_{P_i}^k \oplus C_{P_j}^k \oplus C_{T_{ij}}^k$ can be summarized in Table 1. According to Table 1, we have $C_{P_i}^k \oplus C_{P_j}^k \oplus C_{T_{ij}}^k = 0$. Accordingly, it follows

$$R_{ij}^k = R_{P_{ij}}^k \oplus R_{P_{ji}}^k \oplus C_{T_{ij}}^k$$
$$= \left(G_{P_i}^k \oplus C_{P_i}^k \oplus K_{P_{ij}}^k\right) \oplus \left(G_{P_j}^k \oplus C_{P_j}^k \oplus K_{P_{ji}}^k\right) \oplus C_{T_{ij}}^k$$
$$= \left(G_{P_i}^k \oplus G_{P_j}^k\right) \oplus \left(C_{P_i}^k \oplus C_{P_j}^k \oplus C_{T_{ij}}^k\right)$$
$$= G_{P_i}^k \oplus G_{P_j}^k . \qquad (2)$$

According to formula (2), $R_{ij}^k$ is the XOR value of $G_{P_i}^k$ and $G_{P_j}^k$. If $R_{ij}^k = 0$, it follows $G_{P_i}^k = G_{P_j}^k$; otherwise, $G_{P_i}^k \neq G_{P_j}^k$. It can be concluded that the output of the proposed protocol is correct.

### 3.2.2 Security

Two cases of security need to be considered in this Subsection. One is the attack from an outside eavesdropper, while the other is the attack from a dishonest participant.

**(i) Outside attack**

The outside attack is discussed according to each step of the proposed protocol.

In the proposed protocol, there is not any quantum transmission or classical transmission in Steps 1 and 4. As a result, an outside eavesdropper has no opportunity to launch an attack in these two Steps.

In Step 2, $P_i$ sends $S'_{P_iT}$ to TP. It is apparent that the decoy photon technology [65,66] is used to detect the attacks from an outside eavesdropper in this Step, such as the intercept-resend attack, the measure-resend attack, the entangle-measure attack, *etc*. Actually, this kind of eavesdropping detection technology can be regarded as a variant of that in the BB84 protocol [1], whose unconditional security has been confirmed in Ref.[67]. The effectiveness of decoy photon technology has also been explicitly demonstrated in Refs.[27,28]. In addition, the proposed protocol is automatically immune to the Trojan horse attacks from an outside eavesdropper, such as the invisible photon eavesdropping attack [68] and the delay-photon Trojan horse attack [69], as its quantum transmission is in a single-directional way.

In Step 3, $P_i$ publishes $R_{P_{ij}}$ to TP. An outside eavesdropper may hear $R_{P_{ij}}$. However, she still cannot obtain $G_{P_i}^k$ in this Step, since $G_{P_i}^k$ is encrypted with $K_{P_{ij}}^k$ and $C_{P_i}^k$ which she has no access to.

In Step 5, TP tells $P_i$ and $P_j$ the comparison result of $X^i$ and $X^j$. An outside eavesdropper still cannot get $X^i$ and $X^j$ even though she receives the comparison result.

It can be concluded now that the proposed protocol is immune to the outside attack.

**(ii) Participant attack**

In 2007, Gao *et al.* [70] first pointed out that the attack from a dishonest participant, i.e., the participant attack, is generally more powerful and should be paid more attention to. Three cases of participant attack need to be considered, including the attack from one dishonest user, the collusion attack from two or more dishonest users and the attack from the semi-honest TP.

**Case 1: The attack from one dishonest user**

A dishonest user may try her best to get the other users' secrets. During the comparison between $X^i$ and $X^j$ ($i, j = 1, 2, \ldots, K$, $i \neq j$), there are two situations of participant attack from one dishonest user. One is the participant attack from $P_i$ or $P_j$, while the other is the participant attack from $P_t$, where $t = 1, 2, \ldots, K$ and $t \neq i, j$.

**(a) The participant attack from $P_i$ or $P_j$**

During the comparison between $X^i$ and $X^j$ ($i, j = 1, 2, \ldots, K$, $i \neq j$), the role of $P_i$ is similar to that of $P_j$. Without loss of generality, suppose that $P_i$ is the dishonest user.

Firstly, $P_i$ may try to intercept $S'_{P_jT}$ $P_j$ sends to TP in Step 2. However, she will be caught as an outside eavesdropper since she will inevitably leave her trace, due to having no knowledge about the positions and the preparation basis of decoy photons in $S'_{P_jT}$.

Secondly, $P_i$ may hear $R_{P_{ji}}$ when $P_j$ publishes it to TP in Step 3. However, as for $P_i$, $M_{P_j}^k$ is randomly in one of the two states $\{|0\rangle, |1\rangle\}$. It is equivalent to say that $C_{P_j}^k$ is completely random to $P_i$. Consequently, since $R_{P_{ji}}^k = G_{P_j}^k \oplus C_{P_j}^k \oplus K_{P_{ji}}^k$, $P_i$ cannot know $G_{P_j}^k$ at all.

Thirdly, in Step 5, TP tells $P_i$ and $P_j$ the comparison result of $X^i$ and $X^j$. $P_i$ cannot know $G_{P_j}^k$ either.

**(b) The participant attack from $P_t$ ($t = 1, 2, \ldots, K$ and $t \neq i, j$)**

In the proposed protocol, $K$ users are independent from each other in the sense that they individually prepare quantum sequences and send particles to TP. During the comparison between $X^i$ and $X^j$ ($i, j = 1, 2, \ldots, K$, $i \neq j$), if a dishonest user $P_t$ ($t = 1, 2, \ldots, K$ and $t \neq i, j$) launches an attack, as for $P_i$ and $P_j$, she will essentially act as an outside eavesdropper. According to the analysis on the outside attack in Subsec.3.2.2, the proposed protocol can be immune to the participant attack from $P_t$ ( $t = 1,2,\ldots,K$ and $t \neq i, j$ ) during the comparison between $X^i$ and $X^j$ ( $i, j = 1,2,\ldots,K$, $i \neq j$ ).

**Case 2: The collusion attack from two or more dishonest users**

Two or more dishonest users may collaborate to perform the collusion attack to get the other users' secrets. During the comparison between $X^i$ and $X^j$ ( $i, j = 1,2,\ldots,K$, $i \neq j$ ), there are two situations of this collusion attack. One is that the colluding dishonest users include $P_i$ or $P_j$, while the other is that the colluding dishonest users do not include $P_i$ and $P_j$.

**(a) The colluding dishonest users include $P_i$ or $P_j$**

During the comparison between $X^i$ and $X^j$ ( $i, j = 1,2,\ldots,K$, $i \neq j$ ), the role of $P_i$ is similar to that of $P_j$. Without loss of generality, suppose that in this situation of collusion attack, $P_j$ is a honest user and $P_i$ is one of the dishonest users. Since $K$ users are independent from each other in quantum state preparation and transmission, during the comparison between $X^i$ and $X^j$ ( $i, j = 1,2,\ldots,K$, $i \neq j$ ), each of the dishonest users excluding $P_i$ will essentially play the role of an outside eavesdropper. As a result, if $P_i$ colludes with them, the valuable information available for all colluding dishonest users is equivalent to that under the situation where only $P_i$ is the dishonest user. According to the analysis on Case 1 of participant attack in Subsec.3.2.2, this situation of collusion attack is ineffective.

**(b) The colluding dishonest users do not include $P_i$ and $P_j$**

Since $K$ users are independent from each other in quantum state preparation and transmission, during the comparison between $X^i$ and $X^j$ ( $i, j = 1,2,\ldots,K$, $i \neq j$ ), as for $P_i$ and $P_j$, each of dishonest users excluding $P_i$ and $P_j$ will essentially act as an outside eavesdropper. Even though they collude together, they merely get the same amount of information to that a genuine outside eavesdropper obtains. According to the analysis on the outside attack in Subsec.3.2.2, they cannot obtain anything useful about $X^i$ or $X^j$.

**Case 3: The attack from the semi-honest TP**

The semi-honest TP may try his best to obtain users' secrets without conspiring with either of them.

In Step 3, TP gets $R_{P_{ij}}^k$ from $P_i$. Even though TP can know $C_{P_i}^k$ from his measurement result of $Q_{P_iT_1}^k$, he still cannot deduce $G_{P_i}^k$ from $R_{P_{ij}}^k$, since $G_{P_i}^k$ is encrypted with $K_{P_{ij}}^k$ which he has no knowledge about.

It should be emphasized that TP knows the comparison result of $X^i$ and $X^j$.

## 4 Discussion and conclusion

We further compare the proposed protocol with some current representative two-party QPC protocols, such as the protocols in Refs.[34,36,38-40,44], without considering the eavesdropping check processes, as the eavesdropping check process can be regarded as a standard procedure independent from the working principle of one protocol. The comparison results are summarized in Table 2. It is apparent from Table 2 that each of these protocols relatively has advantages and disadvantages more or less. Undoubtedly, when accomplishing the equality comparison among $K$ users, the proposed protocol exceeds the protocols of Refs.[34,36,38-40,44] in number of times of protocol execution, because arbitrary pair's comparison of equality among $K$ users can be achieved within one execution in the proposed protocol.

Moreover, we compare the proposed protocol with current MQPC protocols in Refs.[51,62-64] also omitting the eavesdropping check processes. The comparison results are summarized in Table 3. Here, the qubit efficiency [12,13,21] is defined as $\eta = \frac{c}{t}$, where $c$ and $t$ are the numbers of compared classical bits and consumed particles, respectively. Note that there are two MQPC protocols proposed in Ref.[63]; accordingly, we represent them with Ref.[63]-A and Ref.[63]-B, respectively. It is apparent from Table 3 that each of the involved protocols relatively has advantages and disadvantages more or less. For example, as the protocol of Ref.[63]-B is based on the $d$-level quantum system, its realization of quantum state preparation and quantum state measurement may be difficult in actual situation. Moreover, it needs unitary operation for users. Fortunately, the proposed protocol does not have these troubles, although its qubit efficiency is not the highest.

On the implementation of the proposed protocol, the technologies for preparing two-photon maximally entangled states as well as storing and measuring single photons are needed. Multi-photon entangled states have been realized in experiment, such as the two-photon entanglement [71], the four-photon entanglement [72], the six-photon entanglement [73], *etc*. The storage of single photons can be realized by optical delays in a fiber [16]. The measurement of single-photon can be accomplished via single-photon detector [74]. Thus, the proposed protocol can be implemented with current technologies.

In summary, in this paper, we summarize eight modes of quantum state preparation and transmission existing in current QPC protocols first. Then, by using the mode of scattered preparation and one-way convergent transmission, we construct a new MQPC protocol with two-particle maximally entangled states, which can accomplish arbitrary pair's comparison of equality among *K* users within one execution, and analyze its output correctness and its security against both the outside attack and the participant attack in detail. The proposed protocol can be implemented with current technologies. It can be concluded that the mode of scattered preparation and one-way convergent transmission of quantum states is beneficial to designing the MQPC protocol which can accomplish arbitrary pair's comparison of equality among *K* users within one execution.

**Acknowledgements**

The authors would like to thank the anonymous reviewers for their valuable comments that help enhancing the quality of this paper. Funding by the National Natural Science Foundation of China (Grant Nos.61402407 and 11375152) is gratefully acknowledged.

**Reference**

[1] Bennett, C.H., Brassard, G.: Quantum cryptography: Public-key distribution and coin tossing. Proc. IEEE Int. Conf. Computers, Systems and Signal Processing, 1984, 175-179
[2] Ekert, A.K.: Quantum cryptography based on Bell's theorem. Phys Rev Lett, 1991, 67(6):661-663
[3] Bennett, C.H., Brassard, G., Mermin, N.D.: Quantum cryptography without Bell theorem. Phys Rev Lett, 1992, 68:557-559
[4] Cabello, A.: Quantum key distribution in the Holevo limit. Phys Rev Lett, 2000, 85:5635
[5] Deng, F.G., Long, G.L.: Controlled order rearrangement encryption for quantum key distribution. Phys Rev A, 2003, 68:042315
[6] Deng, F.G., Long, G.L.: Bidirectional quantum key distribution protocol with practical faint laser pulses. Phys Rev A, 2004, 70:012311
[7] Su, X.L.: Applying Gaussian quantum discord to quantum key distribution. Chin Sci Bull, 2014, 59(11):1083-1090
[8] Zhang, C.M., Song, X.T, Treeviriyanupab, P., et al.: Delayed error verification in quantum key distribution. Chin Sci Bull, 2014, 59(23): 2825-2828
[9] Hillery, M., Buzek, V., Berthiaume, A.: Quantum secret sharing. Phys Rev A, 1999, 59:1829-1834
[10] Karlsson, A., Koashi, M., Imoto, N.: Quantum entanglement for secret sharing and secret splitting. Phys Rev A, 1999, 59:162-168
[11] Xiao, L., Long, G.L., Deng. F.G., Pan, J.W.: Efficient multiparty quantum-secret-sharing schemes. Phys Rev A, 2004, 69:052307
[12] Lin, J., Hwang, T.: An enhancement on Shi et al.'s multiparty quantum secret sharing protocol. Opt Commun, 2011, 284(5):1468-1471
[13] Chen, J.H., Lee, K.C., Hwang, T.: The enhancement of Zhou et al.'s quantum secret sharing protocol. Int J Mod Phy C, 2009, 20(10):1531-1535
[14] Long, G.L., Liu, X.S.: Theoretically efficient high-capacity quantum-key-distribution scheme. Phys Rev A, 2002,65: 032302
[15] Bostrom, K., Felbinger, T.: Deterministic secure direct communication using entanglement. Phys Rev Lett, 2002, 89:187902
[16] Deng, F.G., Long, G.L., Liu, X.S.: Two-step quantum direct communication protocol using the Einstein-Podolsky-Rosen pair block. Phys Rev A, 2003, 68:042317
[17] Deng, F.G., Long, G.L.: Secure direct communication with a quantum one-time pad. Phys Rev A, 2004, 69: 052319
[18] Wang, C., Deng, F.G., Li, Y.S., Liu, X.S., Long, G.L.: Quantum secure direct communication with high-dimension quantum superdense coding. Phys Rev A, 2005, 71:044305
[19] Wang,C., Deng, F.G., Long, G.L.: Multi-step quantum secure direct communication using multi-particle Green-Horne-Zeilinger state. Opt Commun, 2005, 253(1-3): 15-20
[20] Chen, X.B., Wen, Q.Y., Guo, F.Z., Sun, Y., Xu, G., Zhu, F.C.: Controlled quantum secure direct communication with W state. Int J Quant Inform, 2008, 6(4):899-906.
[21] Chong, S.K., Hwang, T.: The enhancement of three-party simultaneous quantum secure direct communication scheme with EPR pairs. Opt Commun, 2011, 284(1):515-518
[22] Liu, D., Chen, J.L., Jiang, W.: High-capacity quantum secure direct communication with single photons in both polarization and spatial-mode degrees of freedom. Int J Theor Phys, 2012, 51:2923-2929
[23] Sun, Z.W, Du, R.G, Long, D.Y.: Quantum secure direct communication with two-photon four-qubit cluster states. Int J Theor Phys, 2012, 51:1946-1952
[24] Ren, B.C., Wei, H.R., Hua, M., Li, T., Deng, F.G.: Photonic spatial Bell-state analysis for robust quantum secure direct communication using quantum dot-cavity systems. Eur Phys J D, 2013, 67:30-37
[25] Zou, X.F, Qiu, D.W.: Three-step semiquantum secure direct communication protocol. Sci China-Phys Mech Astron, 2014, 57(9):1696-1702
[26] Chang, Y., Xu, C.X., Zhang, S.B., et al..: Controlled quantum secure direct communication and authentication protocol based on five-particle cluster state and quantum one-time pad. Chin Sci Bull, 2014, 59(21):2541-2546
[27] Chen, Y., Man, Z.X., Xia, Y.J.: Quantum bidirectional secure direct communication via entanglement swapping. Chin Phys Lett, 2007, 24(1):19
[28] Ye, T.Y., Jiang, L.Z.: Improvement of controlled bidirectional quantum direct communication using a GHZ state. Chin Phys Lett, 2013, 30(4):040305
[29] Jakobi, M., Simon, C., Gisin, N., Bancal, J.D., Branciard, C., Walenta, N., Zbinden, H.: Practical private database queries based on a quantum-key-distribution protocol. Phys Rev A, 2011, 83:022301
[30] Gao, F., Liu, B., Wen, Q.Y., et al.: Flexible quantum private queries based on quantum key distribution. Opt Express, 2012, 20:17411-17420
[31] Gao, F., Liu, B., Huang, W., Wen, Q.Y. : Postprocessing of the oblivious key in quantum private query. IEEE J Sel Top Quant Electron, 2015, 21(3):6600111
[32] Liu, B., Gao, F., Huang, W., Wen, Q.Y.: QKD-based quantum private query without a failure probability. Sci China-Phys Mech Astron, 2015, 58(10):100301
[33] Wei, C.Y., Wang, T.Y., Gao, F.: Practical quantum private query with better performance in resisting joint-measurement

attack. Phys Rev A, 2016, 93: 042318

[34] Yang, Y.G., Wen, Q.Y.: An efficient two-party quantum private comparison protocol with decoy photons and two-photon entanglement. J Phys A : Math Theor, 2009, 42 : 055305
[35] Lo, H.K.: Insecurity of quantum secure computations. Phys Rev A, 1997, 56(2):1154-1162
[36] Chen, X.B., Xu, G., Niu, X.X., Wen, Q.Y., Yang, Y.X.: An efficient protocol for the private comparison of equal information based on the triplet entangled state and single-particle measurement. Opt Commun, 2010,283:1561
[37] Lin, J., Tseng, H.Y., Hwang, T.: Intercept-resend attacks on Chen et al.'s quantum private comparison protocol and the improvements. Opt Commun, 2011,284:2412-2414
[38] Yang, Y.G., Gao, W. F., Wen, Q.Y.: Secure quantum private comparison. Phys Scr, 2009, 80:065002
[39] Liu, W., Wang, Y.B., Cui, W.: Quantum private comparison protocol based on Bell entangled states. Commun Theor Phys,2012,57:583-588
[40] Yang, Y.G., Xia, J., Jia, X., Shi, L., Zhang, H.: New quantum private comparison protocol without entanglement. Int J Quantum Inf, 2012,10:1250065
[41] Chen, X.B., Su, Y., Niu, X.X., Yang, Y.X.: Efficient and feasible quantum private comparison of equality against the collective amplitude damping noise. Quantum Inf Process, 2014,13:101-112
[42] Liu, B., Gao, F., Jia, H.Y., Huang, W., Zhang, W.W., Wen, Q.Y.: Efficient quantum private comparison employing single photons and collective detection. Quantum Inf Process, 2013,12:887-897
[43] Zi, W., Guo, F.Z., Luo, Y., Cao, S.H., Wen, Q.Y.: Quantum private comparison protocol with the random rotation. Int J Theor Phys, 2013,52:3212-3219
[44] Tseng, H.Y., Lin, J., Hwang, T.: New quantum private comparison protocol using EPR pairs. Quantum Inf Process, 2012,11:373-384
[45] Wang, C., Xu, G., Yang, Y.X.: Cryptanalysis and improvements for the quantum private comparison protocol using EPR pairs. Int. J. Quantum Inf, 2013, 11:1350039
[46] Yang, Y.G., Xia, J., Jia, X., Zhang, H.: Comment on quantum private comparison protocols with a semi-honest third party. Quantum Inf Process, 2013, 12:877-885
[47] Zhang, W.W., Zhang, K.J.: Cryptanalysis and improvement of the quantum private comparison protocol with semi-honest third party. Quantum Inf Process, 2013, 12:1981-1990
[48] Li, Y.B., Ma, Y.J., Xu, S.W., Huang, W., Zhang, Y.S.: Quantum private comparison based on phase encoding of single photons. Int J Theor Phys, 2014,53:3191-3200
[49] Liu, X.T., Zhang, B., Wang, J., Tang, C.J., Zhao, J.J.: Differential phase shift quantum private comparison. Quantum Inf Process, 2014,13:71-84
[50] Liu, W., Wang, Y.B.: Quantum private comparison based on GHZ entangled states. Int J Theor Phys, 2012, 51: 3596-3604
[51] Chang, Y.J., Tsai, C.W., Hwang, T.: Multi-user private comparison protocol using GHZ class states. Quantum Inf Process, 2013,12:1077-1088
[52] Li, J., Zhou, H.F., Jia, L., Zhang, T.T.: An efficient protocol for the private comparison of equal information based on four-particle entangled W state and Bell entangled states swapping. Int J Theor Phys, 2014,53(7):2167-2176
[53] Liu, W., Wang, Y.B., Jiang, Z.T.: An efficient protocol for the quantum private comparison of equality with W state. Opt Commun, 2011,284:3160-3163
[54] Zhang, W.W., Li, D., Li, Y.B.: Quantum private comparison protocol with W States. Int J Theor Phys, 2014,53 (5):1723-1729
[55] Ji, Z.X., Ye, T.Y.: Quantum private comparison of equal information based on highly entangled six-qubit genuine state. Commun Theor Phys, 2016, 65:711-715
[56] Sun, Z.W., Long, D.Y.: Quantum private comparison protocol based on cluster states. Int J Theor Phys, 2013,52: 212-218
[57] Liu, W., Wang, Y.B., Jiang, Z.T., Cao, Y.Z.: A protocol for the quantum private comparison of equality with $\chi$ -type state. Int J Theor Phys, 2012,51:69-77
[58] Liu, W., Wang, Y.B., Jiang, Z.T., Cao, Y.Z., Cui, W.: New quantum private comparison protocol using $\chi$ -type state. Int J Theor Phys, 2012, 51:1953-1960
[59] Lin, S., Guo, G.D., Liu, X.F.: Quantum private comparison of equality with $\chi$ -type entangled states. Int J Theor Phys, 2013,52:4185-4194
[60] Ye, T.Y.: Quantum private comparison via cavity QED. Commun Theor Phys, 2017,67(2):147-156.
[61] Ye, T.Y., Ji, Z.X.: Two-party quantum private comparison with five-qubit entangled states.Int J Theor Phys, 2017, 56(5):1517-1529
[62] Liu, W., Wang, Y.B., Wang, X.M.: Multi-party quantum private comparison protocol using $d$ -dimensional basis states without entanglement swapping. Int J Theor Phys, 2014, 53:1085-1091
[63] Wang, Q.L., Sun, H.X., Huang, W.: Multi-party quantum private comparison protocol with $n$ -level entangled states. Quantum Inf Process, 2014, 13:2375-2389
[64] Ye, T.Y.: Multi-party quantum private comparison protocol based on entanglement swapping of Bell entangled states. Commun Theor Phys, 2016, 66 (3) :280-290
[65] Li, C. Y., Zhou, H.Y., Wang, Y., Deng, F.G.: Secure quantum key distribution network with Bell states and local unitary operations. Chin Phys Lett, 2005, 22(5):1049
[66] Li, C.Y., Li, X. H., Deng, F.G., Zhou, P., Liang, Y.J., Zhou, H.Y.: Efficient quantum cryptography network without entanglement and quantum memory. Chin Phys Lett, 2006, 23(11):2896
[67] Shor, P.W., Preskill, J.: Simple proof of security of the BB84 quantum key distribution protocol. Phys Rev Lett, 2000, 85(2):441
[68] Cai, Q.Y.: Eavesdropping on the two-way quantum communication protocols with invisible photons. Phys Lett A, 2006, 351(1):23-25
[69] Gisin, N., Ribordy, G., Tittel, W., Zbinden, H: Quantum cryptography. Rev Mod Phys, 2002,74(1):145
[70] Gao, F., Qin, S.J., Wen, Q.Y., Zhu, F.C.: A simple participant attack on the Bradler-Dusek protocol. Quantum Inf Comput, 2007, 7:329
[71] Kwiat, P.G., Mattle, K., Weinfurter, H., et al.: New high-intensity source of polarization-entangled photon pairs. Phys Rev

Lett, 1995, 75:4337

[72] Weinfurter, H., Zukowski, M.: Four-photon entanglement from down-conversion. Phys Rev A, 2001, 64:010102
[73] Lu, C.Y., Zhou, X.Q., Gühne, O., et al.: Experimental entanglement of six photons in graph states. Nat Phys, 2007, 3:91-95
[74] You, L.X., Shen, X.F., Yang, X.Y.: Single photon response of superconducting nanowire single photon detector. Chin Sci Bull, 2010, 55:441-445

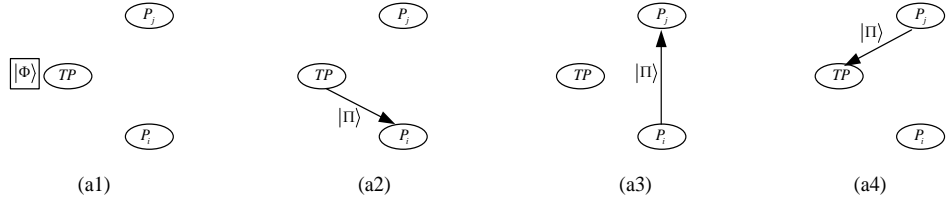

(a)

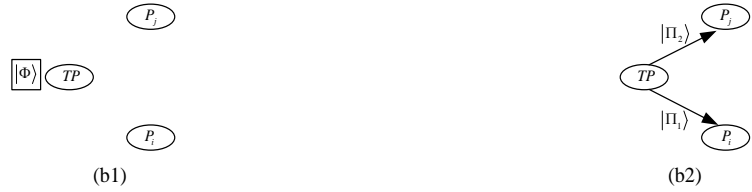

(b)

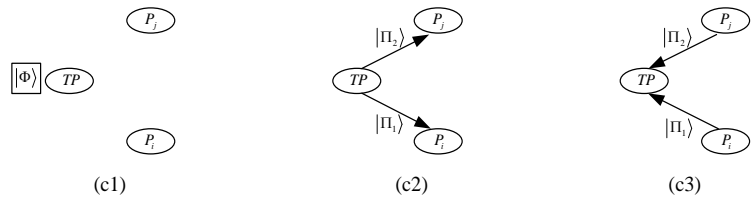

(c)

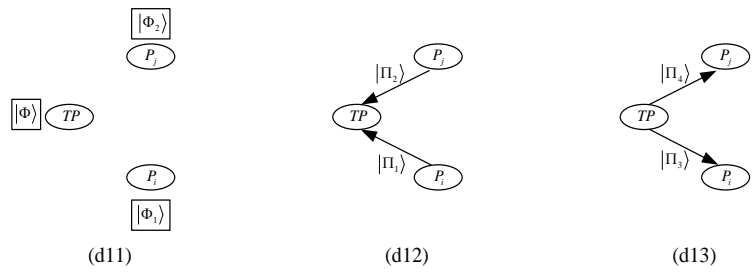

(d1)

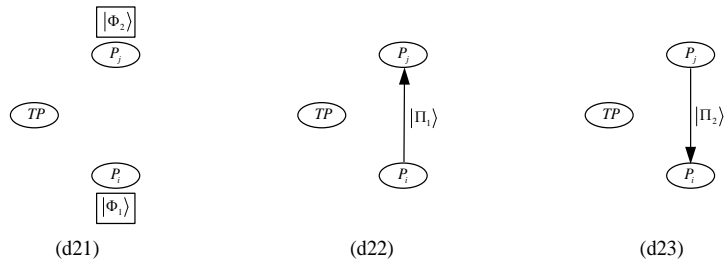

(d2)

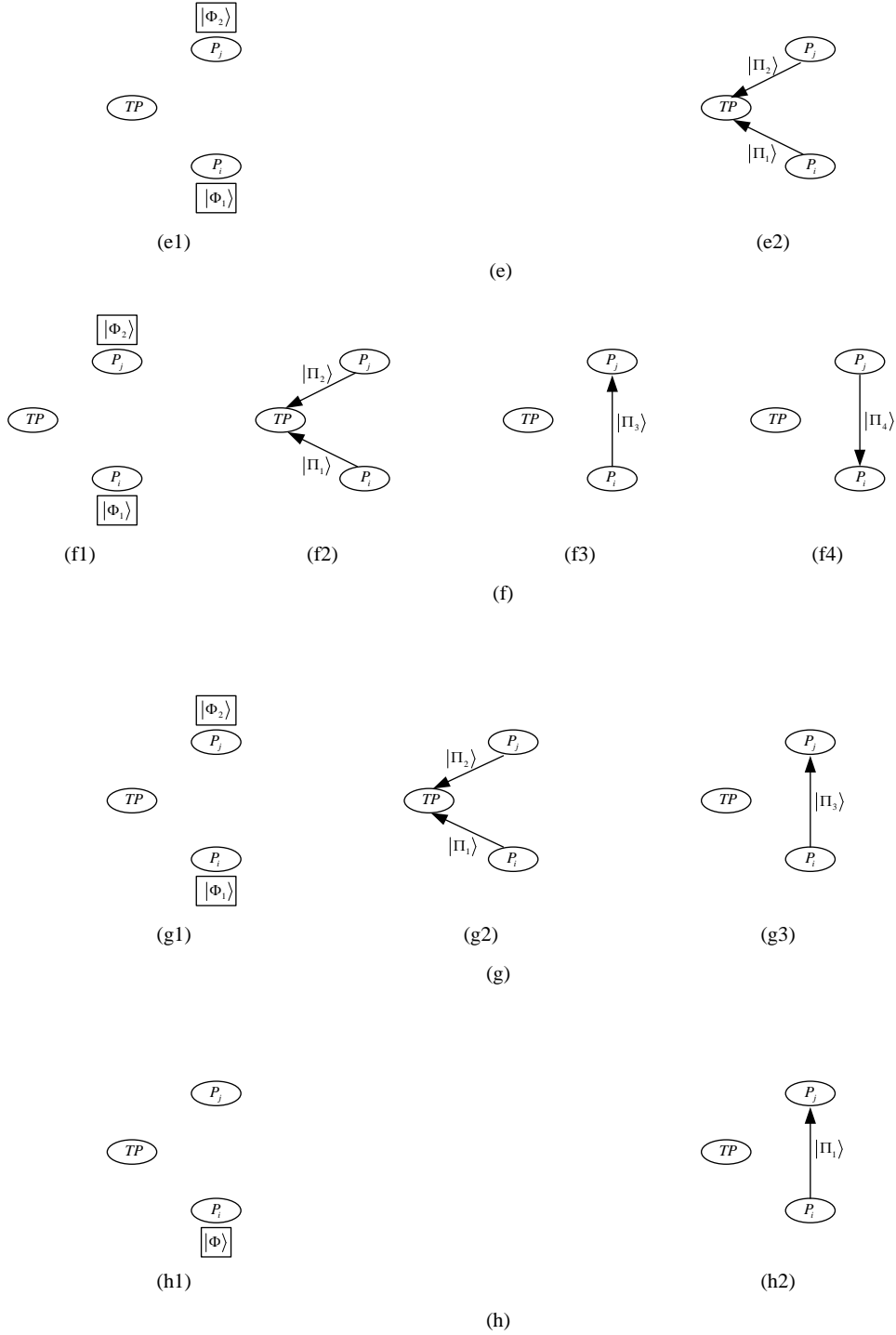

Fig.1 The flow charts of eight modes of quantum state preparation and transmission in current QPC protocols
(a)The mode of centralized preparation and circled transmission. Here, the rectangle with solid lines denotes the quantum state preparation operation, and the solid line with an arrow denotes the quantum state transmission operation; (b)The mode of centralized preparation and one-way divergent transmission; (c)The mode of centralized preparation and bidirectional divergent transmission; (d)The mode of scattered preparation and bidirectional transmission, where d1 is the first case and d2 is the second case; (e)The mode of scattered preparation and one-way convergent transmission; (f)The mode of scattered preparation and hybrid transmission; (g)The mode of scattered preparation and one-way transmission; (h)The mode of centralized preparation and one-way transmission.

Table 1 The relations among essential parameters $M_{P_i}^k$, $C_{P_i}^k$, $M_{P_iT}^k$, $M_{P_j}^k$, $C_{P_j}^k$, $M_{P_jT}^k$, $C_{T_{ij}}^k$ and $C_{P_i}^k \oplus C_{P_j}^k \oplus C_{T_{ij}}^k$

(taking the comparison between $P_i$ and $P_j$ ($i, j = 1, 2, \ldots, K$ and $i \neq j$) for example)

| $M_{P_i}^k$ | $C_{P_i}^k$ | $M_{P_iT}^k$ | $M_{P_j}^k$ | $C_{P_j}^k$ | $M_{P_jT}^k$ | $C_{T_{ij}}^k$ | $C_{P_i}^k \oplus C_{P_j}^k \oplus C_{T_{ij}}^k$ |
|---|---|---|---|---|---|---|---|
| $|0\rangle$ | 0 | $|0\rangle$ | $|0\rangle$ | 0 | $|0\rangle$ | 0 | 0 |
| $|0\rangle$ | 0 | $|0\rangle$ | $|1\rangle$ | 1 | $|1\rangle$ | 1 | 0 |

| | | | | | | | |
|---|---|---|---|---|---|---|---|
| $\lvert 1\rangle$ | 1 | $\lvert 1\rangle$ | $\lvert 0\rangle$ | 0 | $\lvert 0\rangle$ | 1 | 0 |
| $\lvert 1\rangle$ | 1 | $\lvert 1\rangle$ | $\lvert 1\rangle$ | 1 | $\lvert 1\rangle$ | 0 | 0 |

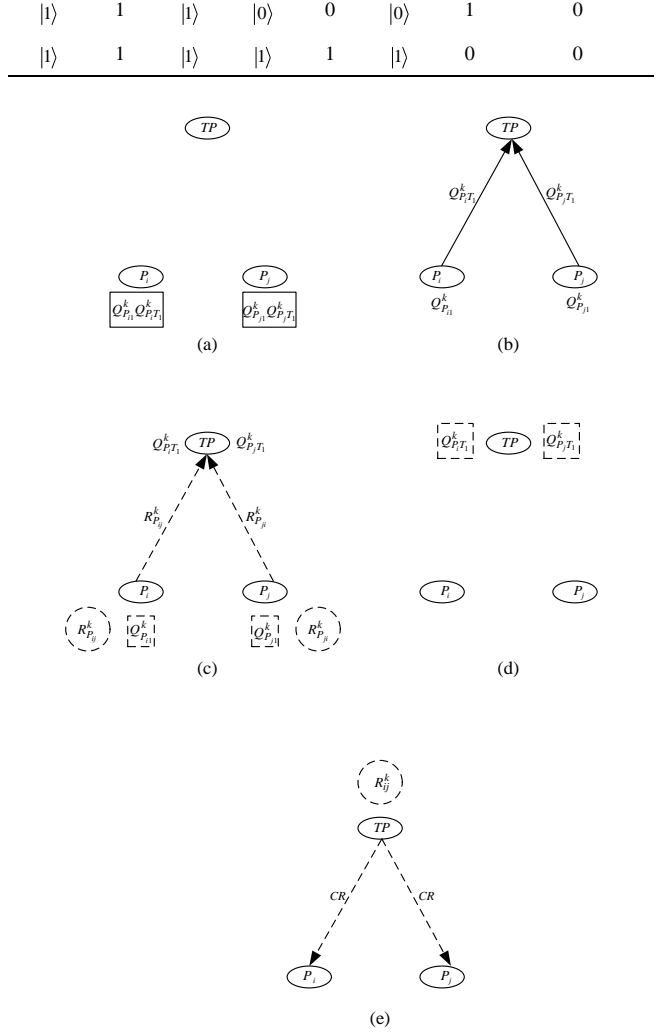

Fig.2  The flow chart of the proposed MQPC protocol
(taking the comparison between $P_i$ and $P_j$ ( $i,j=1,2,\ldots,K$ and $i\neq j$ ) for example)

(a) $P_i$ ( $P_j$ ) prepares quantum state $Q^k_{P_{i1}}Q^k_{P_iT_1}$ ( $Q^k_{P_{j1}}Q^k_{P_jT_1}$ ) as the quantum carrier. Here, the rectangle with solid lines denotes the quantum state preparation operation; (b) $P_i$ ( $P_j$ ) transmits particle $Q^k_{P_iT_1}$ ( $Q^k_{P_jT_1}$ ) to TP, and keeps particle $Q^k_{P_{i1}}$ ( $Q^k_{P_{j1}}$ ) intact. Here, the solid line with an arrow denotes the quantum state transmission operation; (c) $P_i$ ( $P_j$ ) measures particle $Q^k_{P_{i1}}$ ( $Q^k_{P_{j1}}$ ), computes $R^k_{P_{ij}}$ ( $R^k_{P_{ji}}$ ) and sends $R^k_{P_{ij}}$ ( $R^k_{P_{ji}}$ ) to TP. Here, the dotted rectangle, the dotted circle and the dotted line with an arrow denote the quantum measurement operation, the classical computation operation and the classical information transmission operation, respectively; (d) TP measures particles $Q^k_{P_iT_1}$ and $Q^k_{P_jT_1}$ ; (e) TP computes $R^k_{ij}$ and sends $CR$ to $P_i$ and $P_j$ . Here, $CR$ denotes the comparison result.

Table 2.  Comparisons of the proposed MQPC protocols and current representative two-party QPC protocols

| | The protocol of Ref.[34] | The protocol of Ref.[36] | The protocol of Ref.[38] | The protocol of Ref.[39] | The protocol of Ref.[40] | The protocol of Ref.[44] | The proposed protocol |
|---|---|---|---|---|---|---|---|
| Quantum state | Bell state | Triple GHZ state | Single-photon state | Bell state | Single-photon product state | Bell state | Bell state |
| Quantum measurement for TP | Bell basis measurement | Single-photon measurement | Single-photon measurement | Bell basis measurement | No | No | Single-photon measurement |
| Quantum measurement for users | No | Single-photon measurement | No | Bell basis measurement | Single-photon measurement | Single-photon measurement | Single-photon measurement |
| Unitary operation for TP | No | Yes | No | No | No | No | No |
| Unitary operation for users | Yes | No | Yes | No | No | No | No |
| Quantum memory | No | Yes | No | Yes | No | No | Yes |
| Number of times of protocol execution | $(K-1)\sim K(K-1)/2$ | $(K-1)\sim K(K-1)/2$ | $(K-1)\sim K(K-1)/2$ | $(K-1)\sim K(K-1)/2$ | $(K-1)\sim K(K-1)/2$ | $(K-1)\sim K(K-1)/2$ | 1 |

| | The protocol of Ref.[51] | The protocol of Ref.[62] | The protocol of Ref.[63]-A | The protocol of Ref.[63]-B | The protocol of Ref.[64] | The proposed protocol |
|---|---|---|---|---|---|---|
| | | | | | | |

Table 3. Comparisons of the proposed MQPC protocols and current MQPC protocols

| | The protocol of Ref.[51] | The protocol of Ref.[62] | The protocol of Ref.[63]-A | The protocol of Ref.[63]-B | The protocol of Ref.[64] | The proposed protocol |
|---|---|---|---|---|---|---|
| Quantum state | n-particle GHZ class state | d-level n-particle entangled state | d-level n-particle entangled state and d-level two-particle entangled state | d-level two-particle entangled state | Bell state | Bell state |
| Quantum measurement for TP | No | d-level single-photon measurement | d-level single-photon measurement | d-level two-photon collective measurement | Bell basis measurement | Single-photon measurement |
| Quantum measurement for users | Single-photon measurement | No | d-level single-photon measurement | No | Bell basis measurement | Single-photon measurement |
| Unitary operation for TP | No | No | No | No | No | No |
| Unitary operation for users | No | Yes | No | Yes | No | No |
| Quantum memory | No | No | Yes | Yes | Yes | Yes |
| Number of times of protocol execution | 1 (whether the $k^{th}$ group bits from any two parties are equal or not equal) | 1 (whether the $k^{th}$ group bits from K parties are all equal or not all equal) | 1 (whether the $k^{th}$ group bits from K parties are all equal or not all equal) | 1 (whether the $k^{th}$ group bits from K parties are all equal or not all equal) | 1 (whether the $k^{th}$ group bits from any two parties are equal or not equal) | 1 (whether the $k^{th}$ group bits from any two parties are equal or not equal) |
| The mode of quantum state preparation and transmission | Centralized preparation and one-way divergent transmission | Centralized preparation and bidirectional divergent transmission | Centralized preparation and one-way divergent transmission | Centralized preparation and circled transmission | Scattered preparation and bidirectional transmission | Scattered preparation and one-way convergent transmission |
| Quantum technology used | The entanglement correlation among different particles from one quantum entangled state | Quantum fourier transform | Quantum fourier transform | Quantum phase shifting operation | The entanglement swapping of Bell states | The entanglement correlation among different particles from one quantum entangled state |
| Qubit efficiency | $\frac{L}{KL}$ | $\frac{L}{KL}$ | $\frac{L}{3KL}$ | $\frac{L}{2L}$ (without considering the quantum resource used by a QKD protocol) | $\frac{2\lceil L/2 \rceil}{2(K+1)\lceil L/2 \rceil}$ | $\frac{L}{2KL}$ (without considering the quantum resource used by a QKD protocol) |